# Assessing the Impact of Disorganized Background Noise on Timed Stress Task Performance Through Attention Using Machine-Learning Based Eye-Tracking Techniques

**Hubert Huang[1], Jeffrey Huang[2]**

[1]Department of Computer Science, Pacific American School, Zhubei City, Hsinchu 30272 Taiwan
[2]Department of Computer Science, Columbia University, New York, NY 10027 USA

Corresponding author: Hubert Huang (e-mail: huberthuang0930@gmail.com).

**ABSTRACT** Noise pollution has been rising alongside urbanization. Literature shows that disorganized background noise decreases attention. Timed testing, an attention-demanding stress task, has become increasingly important in assessing students' academic performance. However, there is insufficient research on how background noise affects performance in timed stress tasks by impacting attention, which this study aims to address. The paper-based SAT math test under increased time pressure was administered twice: once in silence and once with conversational and traffic background noise. Attention is negatively attributed to increasing blink rate, measured using eye landmarks from dLib's machine-learning facial-detection model. First, the study affirms that background noise detriments attention and performance. Attention, through blink rate, is established as an indicator of stress task performance. Second, the study finds that participants whose blink rates increased due to background noise differed in performance compared to those whose blink rates decreased, possibly correlating with their self-perception of noise's impact on attention. Third, using a case study, the study finds that a student with ADHD had enhanced performance and attention from background noise. Fourth, the study finds that although both groups began with similar blink rates, the group exposed to noise had significantly increased blink rate near the end, indicating that noise reduces attention over time. While schools can generally provide quiet settings for timed stress tasks, the study recommends personalized treatments for students based on how noise affects them. Future research can use different attention indices to consolidate this study's findings or conduct this study with different background noises.

## I. INTRODUCTION

Noise pollution, defined as harmful background noise levels, has inevitably become one of the most significant disturbances to urban life in recent decades. Using Sweden as an example, Bluhm and Eriksson [1] have noted that noise pollution from sources like traffic, talking, railroad, and other areas have all significantly increased between 1997 and 2007. Chepesuik [2] points to urbanization as a possible cause of the increase in noise pollution. Recreational facilities like schools and malls all contribute to the ever-increasing background noise level. With the rise of global urbanization, this increasing trend of noise pollution is set to continue [3]. In light of the increasing plight of background noise, organizations like the WHO have listed noise pollution and background noise as a "serious concern" to address [4]. Singh and Davar [5] contend that these concerns are not unfounded. The study found that background noise causes adverse effects such as "interference with communication, sleepiness, and reduced efficiency," highlighting the serious potential consequences of background noise. A study published in the Macedonian Journal of Health confirms these negative effects [6]. By examining brain activity patterns, the study found that background noise greater than 75 dB significantly inhibits cognitive processes [6]. Instrumental parts of the brain like the frontal and occipital lobes were the most negatively affected by the background noise. Trimmel and Poezl [7] agree with these findings, stating that noise has a significant impact on cognitive processes, especially visual attention.

While levels of background noise have increased, stressful timed tasks like standardized tests or testing in general have also been on the rise. These standardized or non-standardized tests often have strict time limits, which greatly increases the induced stress and anxiety on students [8]. With the increasing importance of standardized tests in academics, there has been increasing pressure on students to perform well in timed testing [9]. In these timed scenarios, no matter mocks or in-class tests, attention becomes an increasingly important factor, as students with ADHD or other attention deficit disorders do significantly worse[10]. Thus, the increased background noise could have important effects on school testing in general.

Although there is abundant research connecting attention and different types of background noise, there is a lack of research correlating background noise directly to performance in tasks, especially stressful timed tasks that are commonly present in schools and the workplace. Thus, this study establishes the connection between background noise and performance in stress tasks.

The study has a few key contributions and novel features:

First, the study finds a significant connection between attention and performance during noise exposure. Specifically, the decrease in attention leads to a decrease in performance when participants are exposed to background noise. This was confirmed with a paired T-test showing a significant decrease in both attention and performance with noise. More importantly, this study solidifies blink rate as a possible attention index indicating performance. Approximately 73 percent of participants had corresponding changes in performance based on their change in blink rate, with those who had a decrease in performance showing an increased blink rate (indicating decreased attention), and vice versa.

Second, the study also finds that there are differing impacts of noise on different students. While most have decreased attention and performance, a smaller proportion of students' performance and attention are positively impacted by the noise. This difference possibly correlates to their self-perception of noise.

Third, using a case study, the study also finds that students with ADHD often have enhanced performance and attention from noise.

Fourth, the study finds that while attention remains constant throughout the control section, it significantly decreases in the latter portions with background noise. While there is no significant change in blink rate from the start to end without noise, there is a significant increase in blink rate near the end of the task compared to the start, indicating decreased attention.

## II. Literature Review

### A. Background Noise and Attention

There is substantial research that establishes the connection between background noise and attention, although opinions are divided regarding background noise's impact on attention. Trimmel and Poezl [7] suggested that disorganized environmental background noises, such as yelping and talking, inhibit cognitive processes, including visual and spatial attention. Attention was measured through electrodes that tracked electric potential changes in the brain, with an increase corresponding to increased attention [7]. The study found that with environmental background noise present, there is a significant decrease in DC potential, showing an inhibitory effect on cognitive processes like visual attention.[7] Likewise, Tristán-Hernández et al. [11] established that there is a positive correlation between the frequency of brain waves, especially theta and beta waves, and attention in college students. The study concluded that environmental noise acts as a significant distractor to college students when studying, significantly decreasing brain wave activity, and thus attention and cognition[11]. Banbury and Berry [12] confirmed the previous findings through surveys with office workers. The study claimed that as much as 99 percent of these workers are affected by at least one type of background noise, such as telephone noises or talking, present in the office. Furthermore, as much as 57 percent of the workers mentioned that their attention has been severely impacted by at least one common office noise [12].

Meanwhile, other common noises like patterned or white noises differ from the previously discussed disorganized background noise. While disorganized noises have no discernable pattern, patterned noises carry specific rhythms. Meanwhile, white noise is a specific type of disorganized background noise that has all audible frequencies as defined in [13]. Patterned or white background noise often leads to the opposite effects on attention, as shown in the following studies. Söderlund et al. [14] refuted the common notion of the negative correlation between visual attention and background noise. The study found that while normal irregular environmental background noises from 62-75 dB expectedly lower the attention span across all the participants, white background noise at around 75 dB increases attention and cognitive performance across participants of all age groups. The study explains that this increase is caused by a phenomenon known as stochastic resonance, where noise up to a certain threshold decreases the noise signals reaching the brain, effectively lowering noise levels [14]. Similarly, Takao et al. [15] found that dotted or patterned background noise, like background white noise, achieves stochastic resonance. The study measured that the attention blink, or the time of temporal limitation that blocks visual attention, is significantly decreased with a moderate amount of patterned background noise, indicating an improvement in attention span [15].

Despite the different effects of patterned noise and environmental noise on attention span, it is important to consider the intrinsically differing effects that background noise may have on each person. Helps et al. [16] emphasized the need for personalized treatment for children of different "levels" of attentiveness. [16] observed that background white noise improves the attentiveness of children with ADHD or students with lower attentiveness levels. If the same amount of background white nose was imposed on children with normal or higher levels of attentiveness, it decreased their attentiveness [16]. Thus, it is not only important to consider the general effects of background noise on the population, but also imperative to give specialized conclusions on different attentive level groups as their reaction to background noise differs.

B.  *Attention and Performance in Stress Tasks*

Previous literature has established a mutual relationship between attention and stress task performance. Chajut and Algom [17] claimed that stressful scenarios or tasks redirect the participant's attention to the stress task at hand to maximize performance, demonstrating the key role that attention plays in the performance of stressful tasks. Momin et al. [18] mention specific factors that induce stress, such as time pressure, often caused through a countdown timer. Moreover, Momin et al. [18] concluded the correlation between heartbeat and stress levels, citing that cardiac activity is a key indicator of stress. The study was conducted through the Montreal Imaging Stress Test (MIST), a cognitive test that induces significant stress in the participants. The study found that as stress levels increased, more attention was allocated to the specific task, highlighting the positive correlation between performance in stress tasks and attention [18].

C.  *Attention Indices*

Attention can be quantified through different metrics. Different indices are used in correspondence with the circumstances of each experiment [19]. When evaluating the effects of background noise on attention or other cognitive performances, studies commonly use EEG-based attention indices, measuring attention based on the brain's electrical signals. Chiang et al. [20] justified using an EEG-based attention index. The study argued that EEG is inherently reflective of brain attention levels. Thus, EEG-based indices can accurately quantify attention in learning environments after decoding the brain waves. However, the study noted that using EEG has limitations because it is difficult to decode the signals, and these signals can be easily affected by other cognitive processes [20]. Kobald et al. [21], published by the Journal of Nature, employed an EEG-based attention index in a study that correlates the specific background noise of MRI scanners and the attention of participants, reinforcing the applicability of the EEG attention index in the context of background noise. Likewise, the aforementioned study by Trimmel and Poezl [7] employs a similar attention index as Kobald et al. [21]. Different from the typical EEG attention index, Trimmel and Poezl demonstrated that the DC potential difference across the brain also measures cognition; the electric potential difference from the brain positively correlates with brain activity and cognition, as an increase in the brain's energy consumption would likely correspond to attention span increase[7].

Other than the EEG attention index, another way to gauge attention is through using surveys on self-reported attention. In the previously mentioned study, Banbury and Berry [12] used surveys to gauge the impact of attention from background noises. The survey asked participants to self-report the extent to which different office noises affect their attention. However, Keith et al. [22] claimed that these "self-mindfulness" surveys that measure attention mostly need to include behaviors that correlate with attention to strengthen the validity of these questionnaires. Thus, this study implied that surveys should not be used in a vacuum, but rather as a supporting metric for supporting and corroborating existing behavioral trends.

An alternate way to gauge attention that has risen in traction is eye tracking, often using machine learning. In a study published in the Journal of Experimental Psychology, Vasilev et al. [23] used eye-tracking methods to correlate background noise and reading speed. Judging from the speed and frequency of eye movements, the study observed that extensive amounts of background noise lead to decreased attentiveness. The decreased attentiveness was evident in the constant eye movement backward to reread, indicating distraction and disruption during reading due to background noise [23]. Moreover, Ettenhoffer et al. [24] justified using eye-based metrics to measure attention. The study evaluated one such attention index, the Bethesda Eye & Attention Measure model. After evaluation, Ettenholfer et al. claim that saccadic metrics, or metrics based on eye movements, hold promising results for delivering accurate quantifications of visual attention [24]. Moreover, Huang et al. [25] employed eye-tracking techniques to measure attention during standardized testing, which is a stressful scenario due to time constraints. This reinforces the applicability of eye tracking as an attention index for both stressful or timed tasks.

D.  *Gap*

The aforementioned research provides clear connections between background noise and attention as well as the correlation between stress and performance. There is extensive research on the correlation between different types of noise and attention, with extensive studies on the possible ways to quantify attention. However, there is a lack of research that addresses the relationship between background environmental noise and performance in timed or stressful tasks, especially using attention as a factor in bridging the gap between the two. Thus, this paper will examine the relationship between disorganized background noise and performance in timed stress tasks using the attention index.

E.  *Hypothesis*

This study hypothesizes that background environmental noise generally decreases attention in timed stressful tasks, which leads to a decrease in performance. However, while attention is generally adversely affected by the noise, there may still be significant differences in the effect of background noise between attentive and inattentive students.

III. **Methodology**

A.  *Overview*

The experiment aims to find the correlation between the performance of timed stress tasks and background environmental noise through the enhancement or deterioration of attention. While there is abundant research on the correlation between background noise and attention, the results often greatly vary due to the types of background noise used. Moreover, background noise's effect on attention is often self-reported, rendering inaccuracies due to the subjective nature of the data. Thus, this study used more general forms of disorganized background noise, such as normal conversational background noise or traffic background noise, while incorporating an eye-tracking attention index to provide a quantifiable and unbiased metric for measuring attention. The quantitative index of attention was used to effectively correlate the effect of background noise on the stress task performance. There is one control group of 30 people and one experimental group of 30 people. The control group was not exposed to noise of any kind, only completing two tests to test the difference in difficulty of the 2 tests. The experimental group did the same 2 tests except the first test was done without noise and the second test was done with noise. This would be done to compare the effects of the background noise on stress task performance and attention.

B.  *Participants*

Since this study focuses on student populations commonly exposed to timed stress tasks, the participants are selected from an international high school where tests are used frequently as a metric for academic performance. As the results of the timed tasks may be affected by the students' inherent skills, the study will use a random cluster sample. The study sampled two class blocks, containing students with a wide range of skill levels. All the test subjects have given consent to participate in the experiment. Before the stress tasks, each participant will be given clear instructions on how the test will be administered and the types of questions on the test. The intention of the tests will not be disclosed to the participants in advance to eliminate the Hawthorne effect [26].

C.  *Survey*

After the participants were gathered, they were told to do a survey. The survey includes basic information on each participant, such as their name, age, gender, and highest math level exposure. These collected data all help the study better understand the participants' demographics, possibly identifying confounding variables that could affect the results of this study. Additionally, there was another question for the participants about the self-perceived effects of background noise on their attention. The question is scored on a scale of 1-5. A score of 3 indicates background noise has no little to no effect on their attention, a score of 1 and 2 indicates that noise either negatively impacts attention or somewhat negatively impacts attention, and a score of 5 and 4 indicates that the noise either positively impacts attention or somewhat positively impacts attention. This question would be used when evaluating the distinctions between people who experience a decrease in attention due to noise and those who experience a decrease in attention. All results of this survey remained anonymous and were not disclosed.

D.  *Timed Stress Task*

As mentioned before, one of the key factors in a scenario is time pressure to complete a task, which fully redirects the participants' attention towards that area. One such stress task is the MIST (Montreal Imaging Stress Test), a timed arithmetic test that induces stress through time pressure [18]. In [18], the MIST, which consistently induced stress in participants, was used to evaluate the connections between visual attention and stress, a topic closely related to the study. Supporting the conclusion in [18], Dedovic et al. [27] found a significant increase in cortisol levels, which corresponds to a significant increase in stress levels, when the MIST was conducted. As previously mentioned, background noise often affects the frontal and occipital lobes. Likewise, the brain's frontal lobes play a key role in numerical problems [28]. By using an arithmetic test like the MIST, this study can further explore the impact that background noise has on the brain. Thus, the study will choose an arithmetic test that carries the same stressors and problem types as the MIST.

Therefore, the timed SAT math test without a calculator, a standardized test taken by most high school students, was chosen. The SAT was chosen over the MIST to address the "practice effect," where participants might improve simply from familiarity with a task [29]. Wesnes and Pincock [30] identified the practice effect as a significant challenge in cognitive experiments, as the number of times practiced becomes a confounding variable. If the study were to use the MIST, it would be possible that they improved the second time with noise since they were more familiar with the questions and the platform. By using the SAT, a test already familiar to most high school students, rather than the more foreign MIST, the study minimized this effect, ensuring that the results more accurately reflect the background noise's effect rather than repeated practice.

FIGURE 1. Sample Paper SAT Math Without a Calculator Multiple Choice Question [31]

If $\frac{x-1}{3} = k$ and $k = 3$, what is the value of $x$?
A) 2
B) 4
C) 9
D) 10

FIGURE 2. Sample Paper SAT Math Without a Calculator Short Answer Question [31]

$x + y = -9$
$x + 2y = -25$

According to the system of equations above, what is the value of $x$?

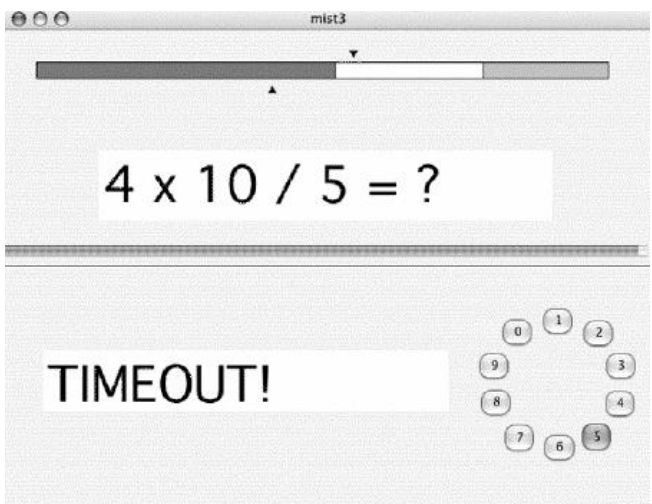

FIGURE 3. Sample Paper SAT Math Without a Calculator Short Answer Question [27]

Comparing Figure 1 and Figure 2 to Figure 3, there are still differences between the SAT Math test and the MIST test. The SAT math test has a lot more words and descriptions, while the MIST test largely comprises arithmetic problems. However, aside from the descriptions, the arithmetic skills employed from these 2 tests are largely similar, still making the SAT Math without a calculator suitable for this occasion.

While the normal test is 25 minutes long, the time was decreased by half to 12.5 minutes to increase the time pressure and stress it had on students. The time of 12.5 minutes was chosen as the time is long enough for students to complete the test but short enough to ensure there was a significant time stressor. Before the study, each participant was instructed on the question types and how to answer them. During the test, each participant will have a timer on their screens, fulfilling the factor of a time stressor. The tests of similar difficulty were chosen and were separately tested using a paired T-test to ensure that they were similar in difficulty.

E. *Background Noise*

In this experiment, the background noise is modified to observe the effect of the task performance. The participants in the experimental group did each test under varying levels of background noise intensity, measured by the dB of the noise. As previous studies have shown, disorganized background noise tends to negatively impact participants' attention, while patterned noise or white may have null or even positive effects on attention. Since this study focuses on background environmental noise, which is closer to disorganized and unrhythmic background noise, it used background noise without significant patterns. The noise file is around 15 minutes in length, corresponding to the length of the test. Following Trimmel and Poezl [7], to amplify the effect that background noise would have, traffic noises and irrelevant conversational noises are used. Using a free audio source Freesound.org, these 2 files were used: "crowd-16.flac" and "BB_0100_approach.mp3" [32] [33]. The first file is a conversational noise and the second is a tunnel traffic noise. It is important to note that the conversational noise does not have discernable conversations, which minimizes the "halfalogue effect" that leads to participants involuntarily paying attention to the conversation in the background [34]. The 2 separate recordings were combined into one audio file at similar volumes. All participants in the experimental group were exposed to this same noise. The noise was played in a 180-degree speaker setup to ensure that the noise was immersive. Moreover, there were also speakers at the 5-degree angle and -5-degree angle on each side of the participant, simulating conversations right beside the person [35]. The experimental trial has a dB of around 80-90 while the control trial has a dB of 40-50.

F. *Attention Index*

As previously mentioned in the literature review, 3 main attention indices were analyzed. Eye-tracking was decided as the main attention index. Although EEG-based indices are common, they were ruled out due to the inconsistencies caused by other psychological signals. Since other phenomena like psychological stress are induced aside from the frontal cortex activity due to arithmetic tests, it is thus probable that the psychological signals would overlap, making it difficult to accurately extract EEG signals correlated with attention. Surveys were ruled out as they required other behavior patterns to support the results of the surveys. This suggests that surveys cannot be used in a vacuum, but only play a supporting role to other indices, making it unsuitable as the main attention index. Thus, eye tracking presented itself as a viable option. Eye tracking is reliable as it can not only stand alone as a separate attention index but also be easily tracked and analyzed, unlike EEG attention indices.

One established attention index using eye tracking is the blink rate. Maffei and Angrilli [36] emphasized the connection between blink rate and attention. The study claims that when participants are more attentive, the blink rate decreases; conversely, a higher blink rate signifies lower attentiveness. Therefore, the blink rate is inversely proportional to attention. Moreover, Maffei and Angrili [36] highlight the use of blink rate as a possible index to study the environmental factors on attention, making it more applicable to a study on background noise. Thus, this study will use blink rate as a quantitative metric for attention.

To track the eyes, OpenCV and Dlib packages will be used in conjunction with Python. A pre-trained facial landmark detection model by Kazemi and Sullivan [37] will be used, which contains 68 landmarks on any given face. The 6 landmarks around each eye will be used for this eye tracking. For the blink frequency, the blink detection will be calculated based on Soukupová and Čech's [38] calculations of Eye Aspect Ratio (EAR).

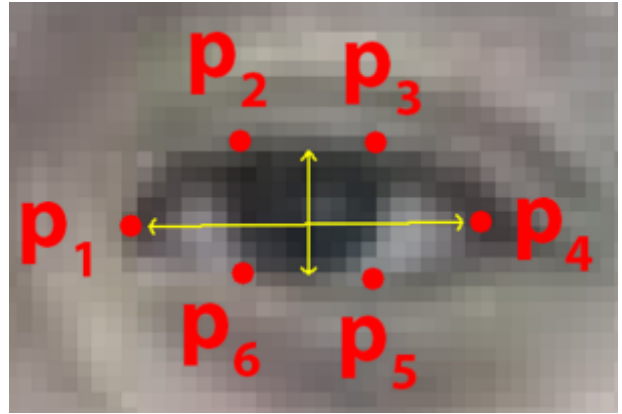

FIGURE 4. **Eye Landmarks used in Eye Aspect Ratio calculation [38]**

$$EAR = \frac{||p_2 - p_6|| + ||p_3 - p_5||}{2||p_1 - p_4||}$$

The EAR is the ratio between the average vertical distances between the top and bottom of the eye or the distance between landmarks 2 and 6 or landmarks 3 and 5, and the horizontal distance, or the distance between landmarks 1 and 4. Thus, the higher the EAR, the wider the eyes are open since the vertical length has increased relative to the horizontal length. Once the EAR passes the lower threshold of 0.2, it will register as a blink, as an EAR of 0.2 means that the eyelids are nearly shut.

The study's code references the code by Rosebrock [39] from PyImageSearch for blink detection with minor changes and addons to fit the purpose of this study. There will be additions like adding and generating the blink rate frequency chart, as well as the addition of downloading the code's frame-by-frame blink rate detection photos for additional analysis.

Despite the eye tracker being consistent for a large number of cases, there still are certain scenarios where blinks are falsely detected.

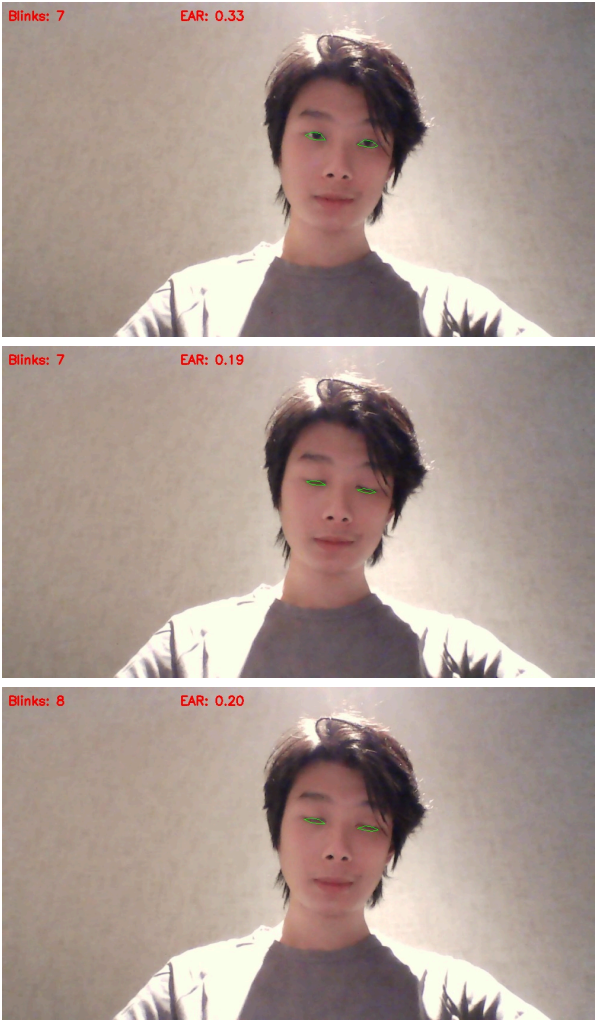

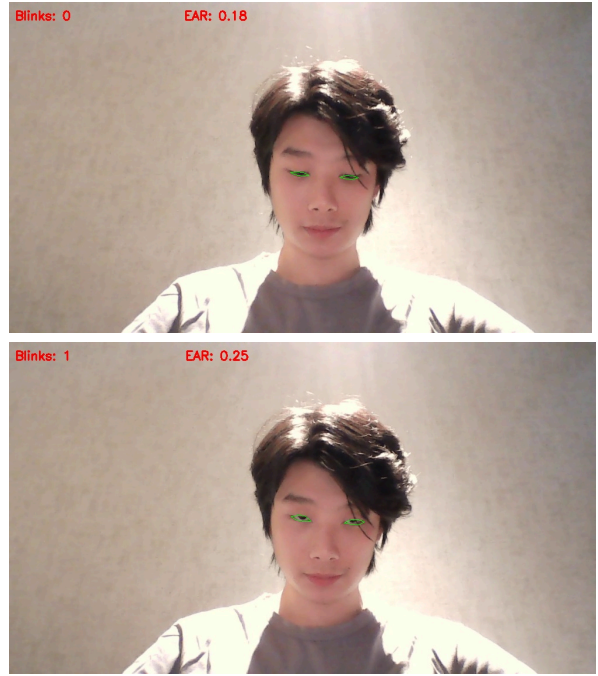

FIGURE 6.  **Sample of a blink cycle where the blink is falsely registered**

However, figure 6 is an example where blinking is falsely registered. In the figure, the person simply first lowers and raises their head. The angle at which the camera captures the face makes the detector falsely detect a blink when the participant does not blink at all. This represents an unsuccessful blink detection where the model falsely recognizes a blink.

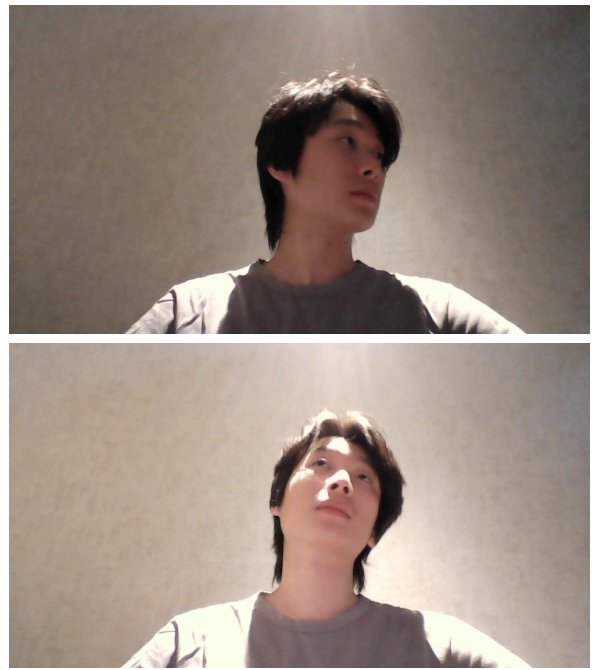

FIGURE 5.  **Sample of a successful blink cycle where the blink is correctly registered**

Figure 5 represents 3 representative snapshots of a successful blink cycle. The person in the first picture in Figure 5 is just about to blink, with his eyes closing in the second photo. As his eyes are closed and his EAR is lower than 0.2 for more than 3 consecutive frames, it registers as a successful blink, as apparent from the blink counter increasing from 7 to 8.

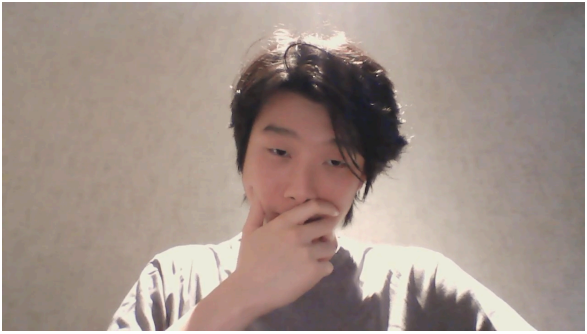

**FIGURE 7.** Sample of situations where the eye cannot be registered

Figure 7 represents scenarios where the facial detection model cannot detect a face, thus no binks can be registered. When a participant looks too much to the left, right, up, or down, the full face is not apparent on camera. Thus, the facial detection model cannot detect all the facial landmarks. Another scenario is where the participant puts a hand on the face, which covers part of it, making the facial detection once again unable to mark the facial landmarks; the blink detector is also incapable of detection during these scenarios.

This study derives a participant's average blink rate by dividing the total number of blinks by the time length of the test. The higher the average blink rate, the lower the attentiveness, and vice versa. The data will be manually adjusted so that these scenarios of misdetection are corrected and times that the facial detection model is unable to detect the facial landmarks will be excluded from the average blink rate calculation.

### G. *Data Analysis*

To analyze the data, this study used a matched pairs T-test for the mean difference. Since the test scores of one singular individual are dependent variables rather than independent variables, the matched pairs T-test would be more suitable as it tests the difference between 2 dependent variables [40]. This study then used the matched pairs T-test to compare the difference in scores between the 2 tests in the control group and the experimental group, as well as the average blink rate.

## IV. Results

### A. *General Demographics Analysis*

TABLE I
DEMOGRAPHICS OF PARTICIPANTS

| Categories | Number of People | Percent of Group (%) |
|---|---|---|
| Age | | |
| 15 Years Old | 4 | 16.67 |
| 16 Years Old | 14 | 46.67 |
| 17 Years Old | 12 | 36.67 |
| Highest Math Level | | |
| Algebra 2 | 3 | 13.33 |
| Precalculus | 13 | 40.00 |
| AP Calculus | 14 | 46.67 |

Table I outlines the population statistics of the test group. All participants signed consent forms for the study, allowing them to exit the experiment at any time if there was discomfort. Since this study uses a random clustered sample, it is expected to find a generally even distribution of math level and age across both the experimental and control groups as shown in the table. Since age and grade levels have a significant impact on students' arithmetic skills, it is important to have a generally even distribution to keep the data normal [41].

As shown in Table I, there is a generally even distribution of participants between ages 16 and 17 with a few outliers of age 15. The case is true for the highest math level grade, as there is a generally even distribution of people whose highest grade level of math is Precalculus or AP Calculus, with a few outliers with Algebra 2. Other important confounding variables are avoided due to the matched-pairs nature of the experiment. Moreover, according to the Central Limit Theorem, sample sizes for the control and experimental groups are 30 people, which ensures that the distribution of the sample data is approximately normal [42]. The test was conducted in such a manner that each observation was independent.

### B. *Test Results and Average Blink Rate Analysis*

Table II compares the test scores and attention indices of the students. Matched Pairs T-test was conducted on both the experimental group with noise and the control group without noise. The recordings of the participants in each of the tests will be analyzed using the aforementioned eye movement and blink detector.

TABLE II
PAIRED T-TEST OF CONTROL TEST SCORES, EXPERIMENTAL TEST SCORES, BLINK RATE DIFFERENCES

| Categories | Mean Difference | Standard Deviation Difference | Degrees of Freedom | T Statistic | One-Sided p-value | Two-Sided p-value |
|---|---|---|---|---|---|---|
| Control Score Difference (Test 2 - Test 1) | 0.3 | 3.7246 | 29 | 0.4412 | 0.3312 | 0.6624 |
| Experimental Score Difference (Test 2 - Test 1) | -1.2333 | 2.3589 | 29 | -2.8637 | 0.003851 | 0.007703 |
| Experimental Blink Difference (Rate 2 - Rate 1) | 2.6172 | 5.5999 | 29 | 2.6172 | 0.007975 | 0.01595 |

(*Note*: Test 1 refers to the score of the "control test" and Test 2 refers to the score of the "experimental test", while blink rate 1 refers to the blink rate without noise and blink rate 2 refers to the blink rate with noise)

The experiment was split into 2 groups, each with 30 people. The control group did the 2 test sets of test questions. As shown in Table II, the p-value of the T-test for the control group is 0.6624, which is much greater than any reasonable significance level. This shows no statistical difference in the mean scores between the 2 tests in the control group. However, for the experimental group, a matched pairs T-test was conducted on the difference in mean of the control-experimental group. This study found a mean test score decrease of 1.233 for the experimental group and a p-value of 0.007703, less than any reasonable significance level. This suggests that there is a significant difference between the test scores of the two groups.

The study also evaluated the difference in attention. The attention was calculated based on the average blink rate with and without background noise. The average blink rate is 11.99 blinks per minute while the average mean blink rate is 14.62 blinks per minute after the noise. A paired T-test was conducted on the average blink rate before and after the noise. The two-sided p-value was calculated to be 0.01596. The p-values are less than the significance level of 0.05, showing a significant difference in the blink rate across the 2 samples.

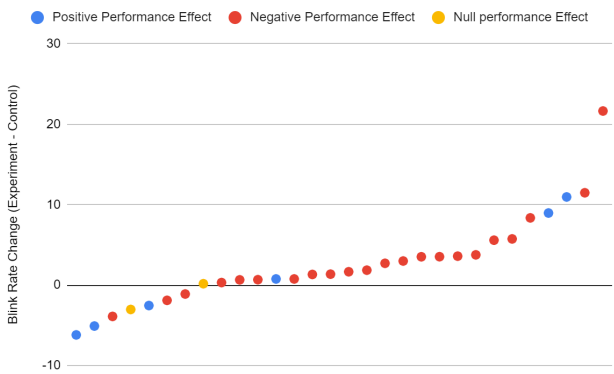

FIGURE 8. **Blink Rate Difference and Performance Effect Chart**

Figure 8 summarizes the statistics presented in Table II. The blink rate was sorted from the most negative change to the most positive change with the blink rate change measured as experimental (with-noise) blink rate minus control (without-noise) blink rate. The color of the dot represents the change in performance: a blue dot means that the score of the participant was increased due to the noise, a red dot means that the score of the participant was decreased due to the noise, and a yellow dot means that there was no change in performance due to the noise.

C. *Sample Blink Rate Frequency Chart Analysis*

Other than the sample analysis of the average blink rate, additional analysis was done on the blink rate. Histograms with time as the horizontal axis and the blinks in one second on the vertical axis. A few representative examples will be placed in this paper for further analysis.

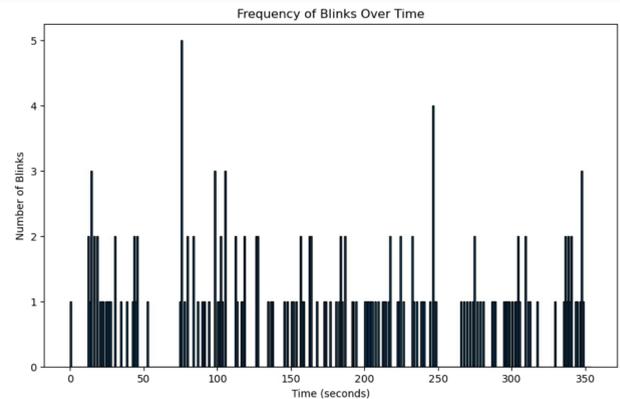

FIGURE 9. **Sample Segment of Blink Frequency Chart for a participant without noise**

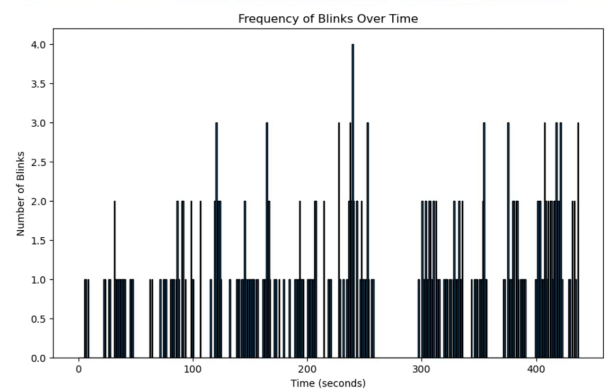

FIGURE 10. Sample Segment of Blink Frequency Chart of the same participant in Figure 4 but with background noise

FIGURE 12. Sample Segment of Blink Frequency Chart of the same participant in Figure 10 but with background noise

For the blink frequency charts, the longer lines represent more than 1 blink in a short period while the smaller lines represent a single blink. As seen in the previous 2 figures, there is a higher frequency of blinks in the latter parts of Figure 10, which is the trial with noise, compared to Figure 9, which was without noise. Figure 9 has a generally even distribution of blinks, indicating that the blink rate was largely consistent throughout the entire interval. However, it is also important to note that this specific participant also experienced a significant blink rate increase of 21 blinks per minute when exposed to background noise. The significant gaps in the middle of Figure 10 and at the start of Figure 9 likely signify that the participant is either covering part of their face or turning away, leading to the facial landmarks detector being unable to recognize the eye landmarks.

On the contrary, Figures 11 and 12 represent a sample participant who experienced a blink rate decrease from the background noise. While this participant could not represent all the other sample participants who experienced a reduction in blink rate due to background noise, this case does serve as a sample. In Figure 11, there is a generally even distribution of blink rates, with occasionally a spike in blink rate. However, similar to Figure 10, in Figure 12, there is an obvious increase in blink rate near the end.

*IV. Initial and Final Average Blink Rate Difference Analysis*

In light of these unexpected findings due to the blink frequency chart, the study thus further analyzes the difference in attention between the beginning and end of the test. The study finds the difference between the average blink rate of the first 1/3 of the video and the last 1/3 of the video. The ratio of 1/3 was chosen as it generally separated the video into the beginning, middle, and end. A section of greater than 1/3 would make for a weak distinction between the middle and end, while a section of less than 1/3 would make for a period that is too short to get a better overview of general trends. A T-test was done for each of the differences, both on the control and experiment, to evaluate whether there is a significant difference in blink rate.

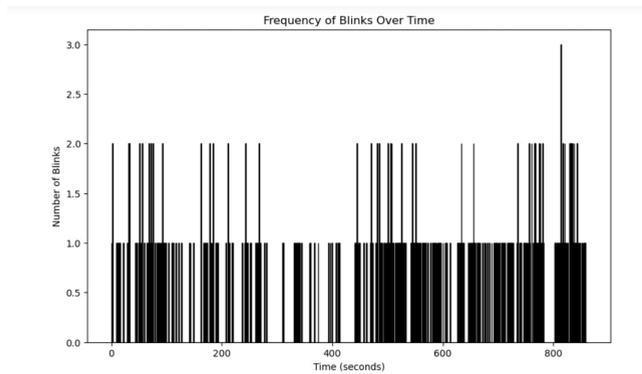

FIGURE 11. Another Sample Segment of the Blink Frequency Chart for a participant without noise

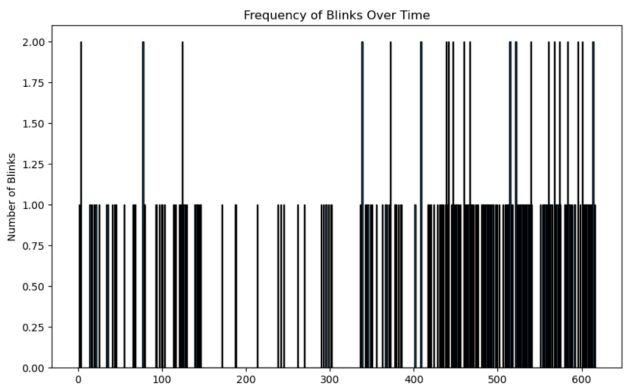

TABLE III
ANALYSIS OF DIFFERENCE IN BLINK RATE AT START AND END

| Categories | Mean Difference | Standard Deviation of Difference | Degrees of Freedom | T Statistic | One-Sided p-value | Two-Sided p-value |
|---|---|---|---|---|---|---|
| Control Score Difference in Blink Rate (End - Start) | 1.5523 | 5.8857 | 29 | 1.4446 | 0.07965 | 0.1593 |
| Experimental Blink Rate Difference (End - Start) | 5.6444 | 7.0502 | 29 | 4.3851 | 0.0000697 | 0.0001394 |

(*Note*: End refers to the average blink rate during the latter 1/3 of the recording and Start refers to the average blink rate during the first 1/3 of the recording)

As apparent in Table III, there is a difference in the change in blink rate throughout the test in the control and experimental groups. In the control group, there was an overall increase in blink rate of 1.5523 blinks per second and a p-value of 0.1593, which is a higher p-value than any reasonable significance level. Thus, there is no significant difference between the start and end blink rate during the control group, which means that attention is generally the same from start to end. On the contrary, for the experimental group, there was an average increase in blink rate of 5.6444. After the one-sample T-test was conducted, the two-sided p-value was 0.0001394. The value is much lower than any reasonable significance level, indicating a significant increase in the blink rate near the end compared to the start.

*V. Detailed Performance Analysis of Different Attention groups*

While there is a general trend of decreasing attention rates, it is also important to acknowledge the limitations of applying general attention trends to the mass public. As mentioned before by Helps et al. [16], there exist different effects of background noise on attentive or inattentive students. While it is not possible to directly determine who is generally attentive and who is generally inattentive without additional attention measurements, this study will still attempt to find distinctions between those who experienced an increase and a decrease in blink rate. Therefore, there will also be an analysis of the self-perceived effects of noise as mentioned in the survey, attempting to make distinctions on the intrinsic qualities between the two groups.

TABLE IV
ANALYSIS OF PARTICIPANTS WITH INCREASED BLINK RATE VS DECREASED BLINK RATE

| Categories | Number of Participants | Average Blink Rate Change | Standard Deviation of Blink Rate Change | Average Score Change | Standard Deviation of Score Change | Average Self-Perceived Noise Effect Scores |
|---|---|---|---|---|---|---|
| Participants With Increased Blink Rate | 23 | 4.4484 | 4.9326 | -1.5217 | 1.9971 | 2.0870 |
| Participants With Decreased Blink Rate | 7 | -3.3994 | 1.6607 | -0.2857 | 3.1997 | 2.7143 |

As apparent from Table IV, there is a clear distinction between the participants who experienced an increase in blink rate compared to those who experienced a decrease in blink rate. Although the general trend of the overall sample of participants is an increase in blink rate corresponding to the background noise, there is still a smaller 7 out of 30 participants that experienced a decrease in blink rate from the background noise. On average, the participants who had a reduction in blink rate had an average reduction of around 3.40 blinks per minute, while those who experienced an increase in blink rate had an average increase of 4.45 blinks per minute. This significant difference is also reflected in the average change in test scores resulting from the noise. Participants who had an increase in blink rate scored on average 1.52 questions less when exposed to noise. On the other hand, participants who had a decrease in blink rate scored only 0.28 questions less when exposed to noise, significantly less than the other group. Moreover, the differences do not just end there. The average self-perceived noise effect scores of the participants with increased blink rate is 2.0870. That score is significantly lower than the average self-perceived noise effect scores of the participants with decreased blink rate, which is 2.7143. These all point to significant differences between these 2 groups of participants.

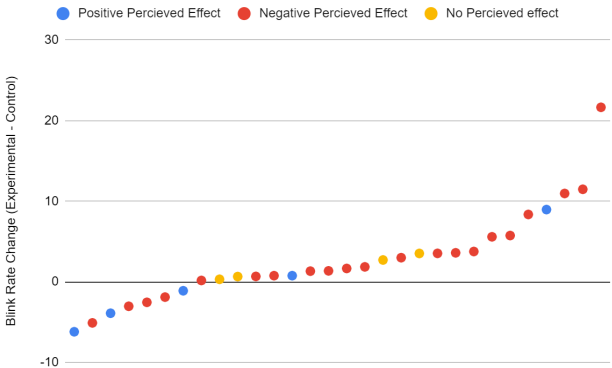

FIGURE 13. **Blink Rate Difference and Self-Perceived Noise Effect Chart**

This graph in Figure 13 summarizes the correlation between the blink rate difference and the self-perceived noise effect scores.

The blink rate like Table 3 is sorted based on the blink rate difference of each participant. Each color marks different groups of participants based on their perceived noise effect scores, with blue dots representing those who viewed that noise had a positive effect on attention, red dots representing those who viewed that noise negatively affected attention, and yellow dots representing those who viewed that noise had little to no effect on themselves.

The following section discusses the interpretations of the data presented.

## V. Discussion
### A. *General Data Analysis*

To analyze the data, it is imperative to first analyze the test scores. From the results, it is apparent that there is a significant difference between the scores with and without noise. With background noise, there was an average score decrease of 1.23. Since it was established using the control group that there is no significant difference between the 2 tests, it thus shows that the background noise leads to a decrease in test scores. This difference can be explained by the measured attention index using the average blink rate.

As mentioned previously, the blink rate increased significantly during the test with background noise. The blink rate increased on average 2.6172 blinks per minute, with a standard deviation of 5.5999. The resulting p-value of 0.01595 thus rejects the null hypothesis that there is no difference in blink rate with or without noises. As mentioned by Maffei and Angrilli [36], this corresponds to a decrease in sustained attention, as an increase in spontaneous blinking inhibits attention toward the environment. This means that the participant's attention was diverted away from the test due to the background noise.

Therefore, combining both the attention index using average blink rate and test scores, this study affirms the first part of the initial hypothesis and concludes that the test scores were likely negatively affected due to reduced attention. This agrees with the conclusion established by Trimmel and Poezl [7]. Despite using different attention indices, both the blink rate-based attention index in this study and the EEG-based attention detection showed a general decrease in attentiveness.

Other than proving the general common conception that background noise negatively affects noise, this T-test also proves the connections between attention, as shown by blink rate, and performance. The study shows that attention measured by blink rate is correlated to performance because there was a significant decrease in attention and a significant decrease in performance when noise was present. This conclusion is also supported by looking at the proportions of participants that had an expected change in performance from their change in attention: decreased attention leads to decreased performance and increased attention leads to increased performance. Out of all 30 participants, 19 out of 30 participants, or 63 percent of the sample, had a decrease in performance with an increase in blink rate. 3 out of the 7 participants with a decrease in blink rate experienced a likewise increase in performance. 2 out of the 30 participants did not have any score change from the change in blink rate. Ultimately, 22 out of the 30 participants or 73 percent of the participants had an increase in performance from increased performance or decreased performance from decreased performance while 2 were simply not affected. This further strengthens the connection between performance and attention.

Moreover, upon further inspection of the blink frequency chart, there were more curious trends. Figures 9 and 10, which come from a sample participant who experienced a significant increase in blink rate when background noise was present, had significant differences. While the blink rate, represented by the density of the lines, was constant mostly throughout Figure 9, which was without noise, the density changed in Figure 10 when noise was present. The blink frequency increased significantly near the end, as seen from the increased density of the lines, while the blink frequency near the start was similar to that in the control group. According to Maffei and Angrili [36], this corresponds to a decrease in attention at times when noise was present while the attention without noise was largely consistent throughout. Moreover, in alternate cases like Figures 11 and 12, where the participant experienced a decrease in blink rate with attention, the same pattern persists. There is a visible increase in the density of the lines near the end of Figure 12, the experimental section, while there is a generally even distribution of lines in Figure 11, which did not have noise. This suggests that, for the participant shown in Figures 11 and 12, noise still plays a role in decreasing attention. The initial attentiveness was high but noise caused attentiveness to deteriorate later on.

The previously described trends were supported by further analysis of the blink rate difference between the end and beginning. In the charts mentioned above, the control group did not have significant changes in attention throughout the test. This is supported by the T-test in Table III. For the control group, there was only an increase of 1.5523 blinks per minute from 13.1 to 14.7 blinks per minute, with a p-value of 0.1593, showing that there is no significant difference between the blink rate of the first third and last third. This means that attentiveness was nearly constant. However, it is different for the experimental group, as there was a high increase of 5.6444 blinks per minute from 14.0 to 19.7 blinks per minute and a low p-value of 0.0001394. This shows that there is a significant increase in blink rate near the end due to noise, which shows a decrease in attention. While the starting blink rates on both the control and experimental were similar, the blink rate at the end of the experimental group was much higher. This presents a new finding as the results seem to suggest that background noise leads to a gradual decay in attention in most cases.

While the trend of decreased attentiveness from background noise holds generally, there are also significant differences among the participants who experienced an increase in blink rate from noise compared to those who experienced a decrease in blink rate. The blink rate increase for participants with increased blink rate is 4.45 blinks per minute while the blink rate decrease for participants with decreased blink rate is 3.40 blinks per minute, which are values with similar magnitudes. This suggests two distinct groups, one group whose attention is positively impacted by noise and one whose attention is negatively impacted by noise, are approximately equally affected by the background noise. The difference in the effect of attention is further reflected by the average score change. Those who experienced an increased blink rate, which indicates a negative effect on attention, had a score decrease of 1.52, which is an even bigger decrease compared to the average score decrease of 1.23. Meanwhile, those who experienced a decreased blink rate, which indicates a positive effect on attention, had only a score decrease of 0.28, which is much less than the overall average score decrease of 1.23. The wide disparity of score change between these 2 groups indicates that there are not only vastly different effects of background noise on different individuals but also that these differences translate to their stress task performance. These distinctions are further strengthened by Figure 8, of the participants that had a decrease in blink rate had a much higher ratio of people who improved in performance due to the noise, as seen from the 3 blue dots below the horizontal axis. This supports the previous distinction that those who had decreased blink rate from the background noise scored better. 50 percent of those who had improved scores due to noise had decreased blink rates, but the people who had decreased blink rates only accounted for a much smaller 23 percent. This means that the probability that a participant had increased performance from the noise is 42.86 percent in the group with decreased blink rate, while only 13.04 percent of participants with increased blink rate increased performance. This significant difference once again solidifies the conclusion that the participants with decreased blink rate and thus performance are likely to perform better.

This agrees with the conclusion of Helps et al. [16], as noise has a widely different impact on each student. While it cannot be established that these students who are positively benefited by the background noise were inattentive without further research, the results do show consistency with the findings. However, another characteristic other than their inherent attentiveness might set these 2 groups apart. In the last column of Table IV, there is a significant distinction between the average perceived noise effect scores. Since a self-perceived noise effect score of 5 represents a positive effect on the noise, 3 represents little to no effect, and 1 represents a negative effect, the higher scores of those who had a decrease in blink rate show that they perceived themselves as inherently more resistant and less likely to get distracted by noise. Meanwhile, the other group that had an increase in blink rate had lower scores, indicating that the participants think that they are more affected by noise. This observation can distinguish between these 2 groups, as people who perceive themselves as more prone to noise may be more affected than those who perceive themselves as less affected by noise. Figure 13 does weakly support this assumption as there is a significantly higher ratio of participants who viewed themselves as positively benefiting from noise in the group with decreased blink rate since 3 out of 7 or 42.9 percent viewed themselves as benefiting from noise contrary to the 9.7 percent in those who had an increase in blink rate.

B.  *Specific Case Study Analysis*

While the previous section explains more general statistics of the participant population, it is also important to consider more individual cases in the participant population. One of the most important case analyses was a participant who had ADHD. According to Söderlund et al. [14], participants with ADHD or other attention deficit disorders may have different reactions than the typical participant. Thus, it becomes important to study this specific case of a student with ADHD and a specific analysis of his reaction to noise and its impact on performance.

The general impact that noise has on performance and attention in this specific case of a student with ADHD is different from the overall trend. The participant's score increased by 4 points when exposed to noise, which is vastly different from the average score decrease of 1.233. This score increase due to noise is also backed up by the blink rate calculations to measure attention. Due to noise, the participant's blink rate decreased by 6.2 blinks per minute from 36.6 blinks per minute in the control group to 30.4 blinks per minute in the experimental group, vastly different from the 2.61 blinks per minute increase due to background noise. Contrary to the rest of the participant sample who had their performance and attention decrease due to noise, this specific sample with ADHD demonstrates the opposite effect as both the participants. This result further consolidates our original findings on the correlation of attention and performance, as the positive increase in attention from the decrease in blink rate correlated with an increase in performance. Moreover, this confirms the conclusion by Söderlund et al. [14], as this study shows that the participants with ADHD benefitted from the background noise that deviates from the rest of the population. However, it is important to note that this conclusion is still based on a singular case study and could be unrepresentative of the entire population.

C. *Implications*

The study proved the initial hypothesis correct. This result can be applied to students in more stressful arithmetic tests. Moreover, with the recent rise in a curriculum based on Science, Technology, Engineering, and Mathematics (STEM), the ability to perform well in testing, especially arithmetic tests, plays an increasingly important role in the education path of students [43]. Since it is now shown that background noise generally reduces attention and also decreases performance in arithmetic tests for the majority of students, schools or individuals can thus apply these findings practically. These results can be used in practical scenarios such as general arithmetic practice or mock tests, where there is a present time constraint but is generally loose on the environmental conditioning. Schools can choose areas with less background noise to conduct arithmetic tests or arithmetic classes to increase student attentiveness. In terms of students, on a general basis, students can choose to use measures such as earplugs or earphones to filter out background noise and to achieve higher attentiveness on these timed arithmetic tasks. However, the study also emphasizes the need for personalized treatments. The study concludes that the vastly different impacts background noise may have on different people; thus the same treatments to decrease noise may not equally benefit all students, since a smaller proportion of students are positively affected by noise. Therefore, it is important to see whether noise is beneficial or detrimental for each student. Lastly, this study also verifies the average blink rate as a valid attention index.

D. *Limitations*

Nevertheless, there still exist limitations. One possible limitation is the degree to which eye blinking accurately represents attention. A confounding variable for using eye blink is the wide range of factors that affect eye blink on a given day. Although Maffei and Angrilli [36] support the notion that eye blinking is strongly correlated with attention, Rodriguez et al. [44] mention the complex nature of blinking. Blinking may not only be impacted by attention, but also by factors such as fatigue or sleepiness, condition of dry eyes, or even just the psychological state of a person [44]. Thus, a portion of the participants' difference in eye movement may be caused by other factors that are not a change in attention. Since this study only uses one attention index, it does not have any other forms of attention indices to support the accuracy of blink rate as a reliable attention index or the findings made by the blink-based attention index. Thus, it would be better to solidify the findings with another attention index. One such index is measuring saccades, or other pupil movements, as an alternative measure of attention to reinforce this study's findings [45].

Another possible limitation is the exclusion of parts of the test. During the average blink rate calculation, there were scenarios as previously mentioned in the methods section when the blink detector cannot accurately register the blinks because the full face was not apparent. These times were excluded to provide a more accurate average blink rate since no blinks could be registered during that period. The times excluded were often when the participant was looking around the room and not on the test. The study was not able to gauge the attentiveness of the participants through blink rate. Especially during the times when the participant is looking out into other directions, the study cannot classify this behavior as deeper thought on the task or mind wandering, a behavior categorized by the participant engaging in other task-unrelated thoughts [46].

E. *Future Directions*

One of the future directions is to directly address limitations. Since this study only uses one attention index, future research could confirm the findings in this paper with similar face detection-based attention indices such as saccades. Zhao et al. [45] exemplified that saccades can also effectively gauge attention, which could reinforce the current attention index. Moreover, other completely different attention indices such as the common EEG-based attention measurement can also be used in conjunction with eye-tracking-based indices to increase the accuracy of the results. Future research could also explore the connections between different types of background noise and stressful arithmetic tests. Since patterned noise and white noise have vastly different effects compared to the disorganized noise used in this study, this could thus differentiate the impacts that different noises may have. Another future direction is to replace the SAT with another timed stress task. Since each timed stress task evokes stress in slightly different ways, future research could study how the effect of background noise changes based on the stressors in a timed stress task.

## CONCLUSION

This study establishes the relationship between disorganized background noise and performance in stressful tasks by using attention as a bridging factor. First, the study affirms the general conception that disorganized background noise decreases both attention and performance. The study also affirms the correlation between blink rate, which represents attention, and performance. Second, the study also establishes that there may be distinct groups that have different reactions to noise, with a smaller group reacting positively to noise as seen from both an increase in performance and attention during noise exposure. This distinction could correlate to the participant's general perception of noise because those who perceive themselves as less negatively affected by noise often experience less decrease or even increase in attention. Third, the study also finds that when exposed to disorganized background noise, attention decreases over time as seen from the increased blink rate near the end of the test compared to the start. Fourth, after analysis of a case study of a participant with ADHD, the study also confirms previous findings on participants with ADHD, as the participant showed a significant increase in both performance and attention after noise. There are various implications to the study. While the findings could be applied generally by decreasing noise during testing scenarios to increase performance, the study also urges for schools to develop personalized treatments for individual students. Based on how a student's attention is affected by noise, schools or teachers can choose how to decrease or increase noise to maximize performance, either by introducing background noise or allowing students to filter out background noise through earplugs or earphones. After this study, future research could be directed at supporting the conclusions made in this study through alternative attention indices such as saccades or EEG indices as well as using alternative types of background noise and stress tasks.

## ACKNOWLEDGMENT

We sincerely thank our advisor Horng-Horng Lin for his committed guidance on our research project. This research would not be possible without his constant support. Moreover, we also thank our teacher Toulouse-Antonin Roy for his assistance with the experimental procedures.

**HUBERT HUANG** was born in Zhubei City, Hsinchu, Taiwan. He is currently a high school student at Pacific American School set to graduate in 2025 and pursue a path in Computer Science. His research interests lie in computer vision.

He is working as an intern in Mabow Technology, developing and training a chatbot API. He has also interned at Yong Cheng Environmental Tech as a data analyst and developed green computing algorithms. His publications include one on the impact of physical exercise on standardized testing and attention.

**JEFFREY HUANG** is currently pursuing a B.S. degree in computer science with a minor in applied mathematics at Columbia University, New York, NY, USA. His research interests include computer vision and natural language processing.

He is working as a Software Engineer Intern at Chimes AI in Taipei, Taiwan, where he contributes to the development of a no-code AI platform for manufacturing. He has previously worked as a Machine Learning Engineer Intern at Teal Omics in New York, NY, and as a Data Analyst Intern at Yong Cheng Environmental in Taoyuan, Taiwan. His publications include research on the impact of physical exercise on standardized testing performance, intellectual property rights for AI-generated music, and reinforcement learning applications.